%% file: main.tex
\documentclass[conference]{IEEEtran}
\usepackage{cite}
\usepackage{amsmath,amssymb,amsfonts}
\usepackage{algorithmic}
\usepackage{graphicx}
\usepackage{textcomp}
\usepackage{xcolor}
\usepackage{url}
\usepackage{hyperref}
\usepackage{longtable}
\usepackage{todonotes}
\usepackage{flushend}
\usepackage{color}

\newcommand\rf[1]{\textcolor{black}{#1}}

\begin{document}

\title{
6GSoft: Software  for Edge-to-Cloud   Continuum
}

\author{\IEEEauthorblockN {Muhammad Azeem Akbar$^1$, Matteo Esposito$^2$, Sami Hyrynsalmi$^1$, Karthikeyan Dinesh Kumar$^3$, Valentina Lenarduzzi$^2$,\\ Xiaozhou Li$^2$, Ali Mehraj$^4$, Tommi Mikkonen$^3$, Sergio Moreschini$^{2,4}$, Niko Mäkitalo$^3$, Markku Oivo$^2$,\\ Anna-Sofia Paavonen$^3$, Risha Parveen$^4$, Kari Smolander$^1$, Ruoyu Su$^2$, Kari Systä$^4$, Davide Taibi$^2$, \\Nan Yang$^1$, Zheying Zhang$^4$, and Muhammad Zohaib$^1$}

\IEEEauthorblockA{\textit{$^1$LUT University}, Lappeenranta-Lahti, Finland ~ \textit{$^2$University of Oulu}, Oulu, Finland}

\IEEEauthorblockA{\textit{$^3$University of Jyväskylä}, Jyväskylä, Finland ~ \textit{$^4$Tampere University}, Tampere, Finland}

\IEEEauthorblockA{azeem.akbar@lut.fi,
matteo.esposito@oulu.fi,
sami.hyrynsalmi@lut.fi,
dinesh.k.karthikeyan@jyu.fi,
valentina.lenarduzzi@oulu.fi, \\
xiaozhou.li@oulu.fi, 
ali.mehraj@tuni.fi, 
tommi.j.mikkonen@jyu.fi, 
sergio.moreschini@oulu.fi, 
niko.k.makitalo@jyu.fi, \\
markku.oivo@oulu.fi, 
anna-sofia.s.paavonen@jyu.fi,
risha.parveen@tuni.fi,
kari.smolander@lut.fi, 
ruoyu.su@oulu.fi, \\
kari.systa@tuni.fi,
davide.taibi@oulu.fi, 
nan.yang@lut.fi,
zheying.zhang@tuni.fi,
muhammad.zohaib@lut.fi}
}

\maketitle

\begin{abstract}
In the era of 6G, developing and managing software requires cutting-edge software engineering (SE) theories and practices tailored for such complexity across a vast number of connected edge devices. Our project aims to lead the development of sustainable methods and energy-efficient orchestration models specifically for edge environments, enhancing architectural support driven by AI for contemporary edge-to-cloud continuum computing. This initiative seeks to position Finland at the forefront of the 6G landscape, focusing on sophisticated edge orchestration and robust software architectures to optimize the performance and scalability of edge networks. Collaborating with leading Finnish universities and companies, the project emphasizes deep industry-academia collaboration and international expertise to address critical challenges in edge orchestration and software architecture, aiming to drive significant advancements in software productivity and market impact.
\end{abstract}


\begin{itemize}

    \item \textbf{Project  Name:} 6G Software for Extremely Distributed and Heterogeneous Massive Networks of Connected Devices 
    \item  \textbf{Acronym:} 6GSoft
    \item \textbf{Project coordinator:} University of Oulu (Davide Taibi)
    \item \textbf{Duration:} May 2023 - April 2026
    \item \textbf{Funded by:} Business Finland 
    \item \textbf{URL:}\url{https://sites.google.com/view/6gsoft}
\end{itemize}
\section{Introduction}

\rf{In recent years,  5G communication technologies have enabled connectivity and innovation \cite{beshley2023emerging}. However, the horizon is already set on the potential of 6G, the next generation of a wave of innovation in wireless communications \cite{96_moussaoui2023divide}. The core components of the 6G infrastructure are heterogeneous networks and a massively distributed nature \cite{dang2020should}. Therefore, 6G requires innovative software development and system integration approaches seamlessly blending emerging technologies like AI, Quantum Computing, and Edge Computing \cite{shen2023five,dang2020should,9114878,9083877}.}

However, current software architectures and development methodologies fail to address the scalability and energy efficiency required for 6G systems \rf{\cite{9585402,9114878,9083877,beshley2023emerging}}. \rf{The ambitious 6G vision and conjectured use cases lead to demands in data rates above 100 Gbps, fraction of millisecond (ms) latencies, and reliability in the order of $10^{-7}$ to $10^{-9}$, which are beyond what is currently supported by 5G \cite{10251878}. Therefore, also Kubernetes\footnote{Kubernetes: \url{https://kubernetes.io}}, the most common orchestrator for cloud and edge computing, needs external help to properly scale and be of any use in this new scenario \cite{10428041}. On the same topic, a recent study showed how Kubernetes was outperformed in IoT use cases leveraging 6G networks \cite{9648317}. } 

The 6GSoft project, a Finnish project funded by Business Finland, aims to investigate sustainable software solutions that are robust, scalable, and energy-efficient. This initiative is not just about keeping pace with technological evolution. Still, it is also driven by the strategic necessity to secure a competitive edge in the global market and avoid past pitfalls experienced by major Finnish corporations in the telecom sector. The goals of the project are manifold:
\begin{itemize}
\item Develop sustainable software development methods tailored for the 6G era, focusing on processes and tools that integrate with heterogeneous systems, including IoT, AI, and quantum computing (QC).
\item Implement energy-aware (EA) orchestration and scalability models that minimize energy consumption while maximizing performance across distributed networks.
\item Foster a business-driven software development model that aligns with emerging 6G technology requirements, ensuring that software architecture and business strategies are cohesively integrated.
\end{itemize}

The outcomes of the 6GSoft project will have a profound impact on both industry and academia. By setting new benchmarks in software development for next-generation edge-to-cloud systems, the project aims to significantly enhance software businesses' productivity, speed, and quality, potentially increasing the market shares by substantial margins. 

The collaboration framework, established through this project, involves four Finnish universities and four local companies, aiming at promoting a robust ecosystem for future technology development in the context of edge-to-cloud and 6G-based systems. Therefore, our paper aims to introduce the 6GSoft project concepts and provide an overview of the literature we are considering as the baseline of our works. 

\textbf{Paper Structure}. Section 2 introduces the project consortium. Section 3 presents the project objectives and the expected outputs. Section 4 presents the current results. Section 5 describes the related works, while Section 6 concludes.

\section{Consortium Description}
The 6GSoft project consortium includes a blend of academic and industrial partners, each bringing unique expertise to address the challenges of 6G software development:

\subsection{Academic Partners}

\noindent \textbf{University of Oulu} - Lead partner with software architecture and engineering expertise for cloud-native systems. 

\noindent \textbf{LUT University} - Focuses on software business ecosystem, digital artifact design, and business models for 6G software.

\noindent \textbf{University of Jyväskylä} - Specializes in orchestration technologies and coordination aspects for enabling the future's 6G Software.

\noindent \textbf{Tampere University} - Focuses on requirements management through AI-assisted techniques and privacy-aware distributed architectures.

\subsection{Industrial steering group}

\noindent \textbf{Bittium Wireless Oy} --- \textbf{Ericsson Oy} --- \textbf{Aidon Oy}  --- \textbf{Wirepas Oy}

\section{Objectives and Expected Outputs}
\rf{6G technology is set to profoundly impact SE by ushering in faster data transmission speeds, ultra-low latency for real-time applications like remote surgery and autonomous vehicles, and the ability to connect many IoT devices simultaneously \cite{su20246g}. Thus, edge computing capabilities should expand and be optimized to support the daunting computational tasks demanded. Moreover, more devices connected to 6G infrastructure will call for new security and privacy challenges toward robust encryption and data protection in SE. On the same vein, EAorchestration will also be a key factor, due 6G's extensive support for device connectivity, where resource efficiency and dynamic allocation policies must be implemented to optimize energy consumption. New AI models such as ``liquid AI'' already exhibit promising results in the new 6G nodes \cite{yang2022liquid}. On the one hand, integrating newly AI-based technologies and paradigms, 6G can potentially optimize network management and security as 6G networks become more pervasive, demanding innovative encryption and cyber defense solutions \cite{taleb20226g,alawadhi2022recent,yang2022liquid}. On the other hand, energy efficiency nowadays is a non-negligible aspect. Ensuring robust security and privacy protections will also be critical.
Finally,  SE development methodologies, e.g., agile, should be tuned to keep up with higher speed, lower latency communication, and faster development cycles for much-needed critical software updates  to real-time infrastructure. \\   Therefore,} the 6GSoft project has four main objectives: (1)To develop sustainable SE practices for the 6G era. (2) To create EAorchestration and scalability models. (3) To provide architectural support for advanced cloud-continuum systems. (4)To formulate business-driven software development models suitable for 6G technologies.

In the following, we introduce our approaches for meeting the objectives with the work conducted in the project's work packages.

\subsection{Sustainable SE practices for the 6G era}

Current SE practices require adaptation to meet the rapid changes demanded by 6G systems. Agile methods and CI/CD adoption are established practices, but their suitability for the 6G era needs assessment. We focus on developing new high-velocity development and deployment practices, encompassing requirements management, metrics, methods, and advanced static and dynamic analysis tools \cite{esposito2024extensive, esposito2023can, esposito2024correlation}. These efforts ensure high software quality while minimizing technical debt and development costs. We also aim to evaluate architectural quality and prevent degradation over time using software metrics, patterns, and Design Rule Spaces \cite{DRS, esposito2024correlation}. Additionally, we are enhancing methods for continuous requirement analysis to mitigate inconsistency and facilitate seamless improvements \cite{esposito2024extensive}.


\subsection{EA orchestration and scalability models}
\rf{As one of the core technologies for 6G, edge computing is an emerging approach that utilizes the computing capabilities of mobile devices\cite{energyware}. To maximize the potential of edge computing, efficient management of containers running in these environments is essential, which requires container orchestration.}
\rf{We are developing EA and highly optimized orchestration and scalability systems. These systems will collect live, reliable data from operational environments (systems and networks) and include diagnostics for analyzing computational evolution. Given that many 6G technologies are still in development, we plan to focus on crucial use cases selected with project partners, such as smart hospitals, real-time AR assistance, future traffic systems, connected vehicles, and generative AI Internet applications to focus on the specific domain or task of which energy should be optimized}. 

From our analysis, we will determine the requirements for orchestration and scalability, guiding the selection of systems and technologies that allow flexible deployment and configuration of future AI-powered applications. To develop orchestration models, we will conduct experiments to assess the practical feasibility of the selected approach and understand, e.g., how state-of-the-art AI models can be harnessed and deployed to the contemporary computing environment. Once the orchestration models are defined, we will explore the potential of artificial intelligence in orchestration and scalability. \rf{Our EAorchestration and scalability models will include mechanisms to adjust resource allocation based on real-time energy consumption data dynamically. Once the orchestration models are defined, we will explore the potential of artificial intelligence in orchestration and scalability, ensuring that energy efficiency is a core consideration.}

\subsection{Architectural support for advanced 6G software}

\rf{Given the complex nature of current software architectures, orchestrating large-scale systems demands substantial resources \cite{wan2023software}. We aim to advance software architectures to facilitate seamless operations for 6G-enable new applications and scenarios. }

\subsection{Business-Driven Development Models}
We are exploring how 6G can transform business models and processes, focusing on creating frameworks that enable companies to leverage 6G capabilities for competitive advantage. We analyze the architectural challenges and produce solutions for 6G software. The solutions are realized as architectural patterns and anti-patterns, standardizable interfaces, and demonstrators that work as a proof-of-concept. 

\section{Summary of Current Contributions}

Regarding \textbf{sustainable SE practices for the 6G era}, we reviewed the literature on AI-assisted requirements engineering \cite{AI4RE}. Our study unveiled a trend of large language model (LLM) based approaches for requirements traceability. We further investigated the effectiveness of these LLM-based approaches to find the traceability links between code commits and requirements. \rf{We are currently expanding the programming languages on which we tested the commits-to-requirement traceability with promising results}. We also investigated the privacy and regulatory challenges involved in the data and machine learning models for 6G infrastructure \cite{kotilainen2023towards,agbese2023examining,Arch,bakhtin2023tools,cerny2022microservice,daubaris2023explainability}. Moreover, we reviewed the state-of-the-art orchestration technologies to understand the feasibility of these technologies for next-generation software in terms of the methods, techniques, and tools reviewed and developed \cite{moreschini2023edge,ESAAM23,Fault,cerny2024static} for this purpose.

Regarding \textbf{EAorchestration and scalability models}, we investigated auction-based orchestration methods for heterogeneous cloud and edge providers \cite{thomsen2023edge}. Our findings include isomorphic implementation methods, particularly WebAssembly \cite{kotilainen2022proposing,kotilainen2023webassembly}. Furthermore, we investigated on multi-layered cloud-native system architecture reconstruction, including testing, pattern detection, and maintenance \cite{cerny2024static,schneider2024comparison,li2023evaluating,cerny2023catalog,smith2023benchmarks,li2023metrics,parker2023visualizing,abdelfattah2023end,cerny2022microservice}

Regarding \textbf{architectural support for advanced 6G software}
We researched a prototype for next-generation software applications involving decentralized deployment in the cloud-edge continuum, leveraging machine learning and language models. Moreover, relevant to this context is our work, as mentioned earlier on prototypes for decentralized, cloud-edge continuum software applications that leverage machine learning and language models \cite{kotilainen2022proposing,kotilainen2023webassembly,su20246g,bakhtin2023tools} and multi-layered cloud-native system architecture reconstruction, including testing, pattern detection, and maintenance \cite{cerny2024static,schneider2024comparison,li2023evaluating,cerny2023catalog,smith2023benchmarks,li2023metrics,parker2023visualizing,abdelfattah2023end,daubaris2023explainability,cerny2024static}. Futhermore, we identified the current research status in SE practices focusing on software process, typical software architecture, orchestration, and offloading methods applied in the context of 6G \cite{su20246g,cerny2023catalog}.
  
Finally, regarding \textbf{business-driven development models}, in the context of 6G software business ecosystem modeling, we investigated bridging the gap between 6G technology and software business, with emphasis on the critical role of the business perspective for 6G commercialization \cite{yang2023exploring,akbar2024role,akbar20246g}. Moreover, we conceptualize an analysis of the 6G ecosystem pioneering research to understand the components, including related concepts, sub-concepts, antecedents, and consequences of the 6G ecosystem. In this context are also the above-mentioned contributions on privacy and regulatory challenges involved in data and machine learning for 6G infrastructure \cite{kotilainen2023towards,agbese2023examining}.

Moreover, we studied the potential role of QC in shaping the future of 6G technology. We discussed the issues of QC for 6G and identified 15 critical applications, which include quantum machine learning optimization, quantum positioning, and enhanced signal processing \cite{akbar2024role}. These applications show the promise of quantum technologies and the vision of a quantum internet in enhancing networks' performance, security, and efficiency. Altogether, these works offer a systematic approach to incorporating quantum technologies into 6G networks with new perspectives. In another study \cite{akbar20246g}, we proposed a success probability prediction model for secure quantum communication in 6G technology employing QKD as a security boosting mechanism. This model incorporates predictive analysis for the probability of secure communication channels to address the critical cyber security requirements for 6G. 

\section{Impact and Threats}
\rf{A possible threat to the validity of our preliminary results on sustainable SE practices for the 6G era lies in the evolving and speculative nature of 6G technology \cite{shen2023five,dang2020should,9114878,9083877}. The reliance on current literature and prototypes may not fully capture future complexities and requirements. Moreover, recent studies have investigated the issues in integrating LLMs and QC and their impact on scalability, security, and regulatory compliance uncertainties \cite{esposito2024classi, sabzevari2024qcshqd, esposito2024beyond, esposito2024leveraging}. Our collective effort will try to address the daunting issue of future 6G development, adapting our models and approaches as the technology progresses.}

\rf{We conjecture our collective efforts will impact the development of new 6G-enabled technologies as well as real-time mission-critical applications, such as remote surgery, and also allow massive IoT connectivity while enabling a structured integration of advanced AI technologies in the edge as well the core of the networks.}

\rf{Finally, despite its broad scope, the 6GSoft project has a clear research plan that will systematically address sustainable engineering, EA models, robust architectures, and business-driven development, ensuring comprehensive advancements in 6G technology. By continuously aligning our research outcomes to the latest advances of the 6G, our project will cross-fertilize ideas between partners, thus improving cross-contributions between academia and industry and significantly contributing to the state of the art. }

\section{Related Work}
\input{RelatedWorks}

\section{Conclusion}
We presented the objectives and current achievements of the 6GSoft project, a collaboration between four leading Finnish universities supported by renowned companies. Our project aims to significantly impact the future of SE for cognitive and cloud continuum in forthcoming 6G environments, integrating sustainability and energy efficiency into the core of 6G software development. We anticipate that the outcome of these joint efforts will be published in top-tier venues within SE and related fields. The partner companies will validate and employ our research effort in their industrial environment. Our project will significantly enhance the close academic collaboration between Finnish universities and industries, paving the way for future global achievements.

\bibliographystyle{IEEEtran}
\bibliography{main}

\end{document}

%% file: RelatedWorks.tex

\subsection{Software Engineering for 6G}

\rf{Recently, studies regarding software engineering practices for 6G, software process, architecture, orchestration and offloading methods, and business-driven software development have been the main topics for 6G software engineering \cite{su20246g}. Since 2022, the number of published papers has increased sharply in this domain. Amongst these publications, software architecture is the area that has received the most attention in 6G Software Engineering. Other covered topics include software tequirements~\cite{SP3}, SE models and methods~\cite{SP7}, software quality~\cite{SP1}, software orchestration~\cite{SP5}, offloading~\cite{SP1} and energy aware~\cite{SP4}. Therein, the software process, architecture, and orchestration publications are still limited, while business-driven software development is nonexistent.}

Al-Hammadi et al.~\cite{SP2} implement a collaborative offloading method among MEC servers based on the edge server’s resources and neighbors’ status to alleviate network congestion and formulate a hierarchy SDN-powered MEC network framework comprising three tiers in the method. Habibi et al.~\cite{SP7} provide guidelines on how the novel software building blocks can be integrated and deployed as part of a DevOps workflow and propose an \rf{M\&O} framework for 6G simultaneously. Alotaibi and Barnawi~\cite{SP11} propose IDSoft, a novel software solution that resides across the network infrastructure and leverages 6G enabling technologies. Shukla et al.~\cite{SP12} propose 6G-SDI, an \rf{SDN}-based green communication method for 6G-enabled Internet of Things (IoT) to control real-time actuation and flow-table configuration. Abdulqadder and Zhou~\cite{SP13} tackle the issues, such as security, quality of service (QoS), and resource consumption issues, through related network slicing and load balancing methods in SDN/NFV-assisted 6G environments. In addition, except for the first two papers that propose the framework in the methods, three other papers present the framework separately. Meenakshi et al.~\cite{SP4} use a framework constructing a design of systematic Wireless Inventory trackers (WIT) using heterogeneous IoT(HIoT) networks over 6G Computing and MEC/SDWAN for improving the prolonged lifetime and low energy consumption for efficient communication in 6G network. Manogaran et al.~\cite{SP5} propose the service virtualization and flow management framework (SVFMF) for the reliable utilization of resources in the 6G-cloud environment. Janbi et al.~\cite{SP17} propose a framework for Distributed AI as a Service (DAIaaS) provisioning for Internet of Everything (IoE) and 6G environments.

Regarding 6G architectures and platforms, Bojović et al.~\cite{SP8} designed a multi-slice architecture to develop highly flexible dynamic queue management software and moved it entirely to the application layer. Alonso-Lupez et al.~\cite{SP9} propose an architecture enabling privacy-aware slicing and security service orchestration. Moreover, Alotaibi and Barnawi~\cite{SP11} propose HFL (hierarchical FL) architecture with an additional offloading mechanism to enhance and evaluate performance in terms of relevant FL aspects, such as accuracy, communication efficiency, and convergence. Tao et al.~\cite{SP18} propose a novel software-defined DTN architecture with digital twin function virtualization (DTFV) for adaptive 6G service response. On the other hand, for the results of "Platform," Ateya et al.~\cite{SP1} develop a MEC platform and introduce a seamless migration for complex 5G/6G tasks. Furthermore, Katiyar et al.~\cite{SP3} develop an IoT platform and focus on its middleware layer connection with other layers. Then Cao et al.~\cite{SP10} propose C-ITS, a Cooperative intelligent transport system that allows softened resource management and allocation in 6G networks with autonomy and smart sensing. Unlike the previous, Kamruzzaman and Alruwaili~\cite{SP14} propose a system and use an optimizing computer vision with AI-enabled technology (OCV-AI model).

Regarding algorithms, Ajibola et al.~\cite{SP6} propose a heuristic for energy-efficient and delay-aware placement (HEEDAP) of applications in fog networks (an algorithm). Cao et al.~\cite{SP15} propose an efficient algorithm, labeled TailoredSlice-6G, to implement the tailored resource allocation of slices in 6G networks. Ye et.al~\cite{SP16} propose a heuristic decoupled SFC orchestration algorithm (HDSFCO) with low complexity to minimize the overall resource costs, where fully consider the time evolution characteristics of dynamic network topology.

\subsection{AI assisted requirements tracing}

\rf{In the era of 6G, collaboration between distributed and heterogeneous stakeholders grows where requirements tracing is critical to the success of software projects. Research on requirements traceability primarily addresses traceability link generation \cite{zhao2021natural,lyu2023systematic,xu2023systematic,corral2022building, sonbol2022use,tao2024magis}, traceability link recovery \cite{zamani2021machine,zhao2021natural,lyu2023systematic,sofian2022systematic,hou2024large}, and bug localization \cite{lyu2023systematic,hou2024large}. }

Many AI methods have been proposed to recover traceability links to achieve reliable results, especially in large, complex software projects with vast repositories, diverse software artifacts, and extensive codebases. Lyu et al.\cite{lyu2023systematic} reviewed 40 primary studies from 2011 to 2022, focusing on using Information Retrieval and Machine Learning techniques to generate and recover traceability links. These studies addressed challenges such as accuracy, effort, support, information, and trustworthiness in requirements traceability. Notably, 50\% of these studies concentrated on traceability link recovery, while 30\% focused on link generation. This finding aligns with \cite{xu2023systematic}, which reviewed 53 primary studies from 2012 to 2021, noting that 26 studies were dedicated to requirements traceability. Other secondary studies on AI for requirements engineering such as \cite{zamani2021machine,zhao2021natural,sofian2022systematic,corral2022building,sonbol2022use,aguilar2020systematic,hou2024large} also emphasized the importance of requirements traceability in research.
 
Before 2018, requirements tracing primarily relied on IR techniques, using methods like TF-IDF, VSM, and Random Forests to automate link recovery between issue reports and commits. Bug localization utilized algorithms such as SZZ, while LSI and LDA improved trace link accuracy. After 2018, research evolved to hybrid techniques, integrating ML, NLP, and other AI methods. Deep learning approaches, including GRUs and LSTMs, enhanced issue-to-code mapping and bug-source identification. Recent studies \cite{lin2021traceability,guo2017semantically,borg2017traceability} highlight large language models (LLMs) like BERT and neural networks for better requirements management, as seen in \cite{hou2024large} TRACEFUN framework, which leverages unlabeled data to improve traceability link recovery.



 \subsection{Energy-aware orchestration and scalability models}

The feasibility of container orchestration technologies for edge computing environments has been analyzed based on resource allocation efficiency, such as memory and CPU utilization. Bahy et al. \cite{bahyResourceUtilizationComparison2023} found Nomad uses CPU and memory efficiently, while K3s excels in storage usage. Telenyk et al. \cite{telenykComparisonKubernetesKubernetesCompatible2021} reported Kubernetes outperforms K3s and Microk8s in memory and CPU utilization, but K3s performs better in storage. Bohm and Wirtz \cite{bohmProfilingLightweightContainer} and Koziolek et al. \cite{koziolekLightweightKubernetesDistributions2023} provided similar findings, with K0s leading in CPU tests and K3s and MicroShift in memory tests.

For energy efficiency, Deep Q-Learning (DQN) \cite{energyware2} optimizes by offloading data, reducing energy use, and shortening service workflow completion times. The Nondominated Sorting Genetic Optimization Algorithm \cite{energyware3} manages to offload complex tasks efficiently in terms of time and energy. The improved Genetic Algorithm (GEA) \cite{energyware4} enhances mobile edge computing performance by optimizing energy consumption and delay for low-latency, energy-efficient, and reliable task execution.

The Nondominated Sorting Genetic Optimization Algorithm \cite{energyware3} addresses energy and resource-aware computation offloading in edge environments. Given the limited resources of heterogeneous edge servers, this optimization algorithm effectively manages offloading complex tasks. It is efficient regarding time and energy consumption, optimizing time, energy, and cloudlet resource utilization.  
The improved Genetic Algorithm (GEA) \cite{energyware4} algorithm offers a significant improvement in \rf{MEC} performance by optimizing energy consumption and delay, thereby supporting low-latency, energy-efficient, and reliable task execution in mobile edge environments.

\subsection{Software Architecture for 6G Era}
\rf{In the 6G era, the need to use architecture definition language has become more critical due to next-generation networks' increased complexity and interoperability requirements \cite{dang2020should,uusitalo2021hexa,lv2020software}. Architecture Description Languages (ADLs) express a system architecture clearly and precisely, thus modeling intervening artifacts of different technologies that ensure seamless and smooth integration, enhanced scalability of mechanisms, and robust security frameworks \cite{ClementsPaul1996ASoA}. Providing a clear blueprint of this type of architecture, ADL helps streamline development processes, cutting down errors, speeding up the delivery of innovative services into reality in 6G, and driving the evolution toward an efficient and reliable communication infrastructure.
Therefore, }we explored the features and functionality of existing ADLs \cite{ClementsPaul1996ASoA}. Unlike modeling languages such as UML, which capture behavior and structure, ADLs primarily focus on structural aspects, emphasizing component representation and interactions \cite{ClementsPaul1996ASoA, MISHRA200759}. This focus allows ADLs to offer a high level of abstraction, facilitating intellectual control over system architecture design and evaluation \cite{haider2018ali}.

The evolution of ADLs can be divided into three periods \cite{EgyhazyCsaba2007Cofa}: \textbf{Theory}. During this phase, research focused on formal approaches to characterize components, behavior, and connections. Notations were developed to define structures. \textbf{First geeneration}. These ADLs primarily focused on modeling architecture elements. \textbf{Second-Generation}: Derived from first-generation architectures, these ADLs emphasize handling dynamic architectures.

UniCon \cite{shaw1995abstractions}, Wright \cite{allen1997formal}, ACME \cite{garlan2010acme}, Rapide \cite{luckham1995specification}, SRI SADL \cite{moriconi1997introduction}, C2 ADL \cite{medvidovic1996formal}, and MethaH \cite{binns1996domain} are considered first-generation and AADL \cite{feiler2006architecture}, xADL \cite{DashofyEric2005Acaf}, ADML \cite{WangZhuxiao2012AADL}, and ArchWare \cite{1604766} are considered second-generation ADLs \cite{EgyhazyCsaba2007Cofa}. As mentioned above, the ADLs were developed in the mid to late 1990s and early 2000s. 

While \rf{ADLs} have gained acceptance in the research community for describing system designs, their adoption in industry remains relatively low. Notable exceptions exist, particularly in the embedded systems domain \cite{haider2018ali}, where ADLs like AADL \cite{feiler2006architecture} and EAST-ADL \cite{CuenotPhilippe2010TEad} have found success and are widely used in avionics and automotive industries. However, ADLs developed in academia, such as ACME, UniCon, and xADL, have not achieved similar industry adoption \cite{haider2018ali}.

Most currently used \rf{ADLs} are either tailored to specific application areas or focus on particular system components, such as system structure. Unfortunately, these ADLs have not scaled well over time \cite{BashroushRabih2006TMFA}. Several existing drawbacks contribute to the limited adoption of ADLs in the industry:

The challenges associated with Architectural Description Languages (ADLs) include \textbf{the lack of robust Computer Aided Software Engineering (CASE) tools} to support their adoption \cite{BashroushRabih2006TMFA, BashroushR.2008AAEA}, \textbf{insufficient mechanisms for enforcing consistent communication styles} among stakeholders, \textbf{limitations in presenting multiple views} of architectures \cite{BashroushRabih2006TMFA, BashroushR.2008AAEA}, \textbf{inadequate support for tracing requirements} \cite{haider2018ali}, \textbf{restrictive syntax} that limits flexibility in real-world applications \cite{haider2018ali}, \textbf{challenges in scaling} to support large applications \cite{haider2018ali}, and \textbf{inefficient artifact reusability} \cite{haider2018ali}.

These challenges collectively contribute to the industry's cautious approach toward adopting ADLs.

While we haven't encountered any specific cases where ADL-type approaches have been used to express and analyze data, ethical, and other concerns, we believe a similar approach could be valuable in constructing and analyzing ethical systems related to data and AI. Hence, it contributes to 6G adoption and development.

\subsection{6G Software Business Ecosystem}
\rf{Research on the 6G ecosystem has increased substantially over the past five years, capturing the interest of both academics and industry \cite{katz20186genesis}.} This heightened attention arises from the widespread anticipation that 6G will serve as a pivotal game changer \cite{03_bhat2023ecosystem} compared to its predecessors. Although 6G technologies and standardization are still in their developmental stages and are expected to be commercialized in 2030 \cite{03_bhat2023ecosystem}, they have attracted significant interest due to their opportunities for ecosystem-level business innovation. Such innovation in designing digital services necessitates collaboration across multiple perspectives, disciplines, and stakeholders \cite{ahokangas2022envisioning}. Unfortunately, despite the extensive attention directed towards the 6G ecosystem and its concept variants, audiences often struggle to understand its antecedents and precisely grasp its essence due to a lack of concept-level research work.

Many studies on 6G ecosystems adopt the concept without a clear definition. For example, \cite{12_azari2023thz} discusses unmanned aerial vehicles in the 6G ecosystem without defining it, and \cite{liao2016ict} explores digital service design similarly. This misuse risks turning the "6G ecosystem" into a vague term \cite{hyrynsalmi2019ecosystem}. For instance, \cite{kovtun2023formalization} proposes a design framework for 4G, 5G, and 6G ecosystems without defining 6G. Additionally, \cite{96_moussaoui2023divide} uses related terms like "mobile network ecosystem" and "beyond 5G ecosystem" interchangeably without clear definitions.

%% file: main.bbl
\begin{thebibliography}{100}
\providecommand{\url}[1]{#1}
\csname url@samestyle\endcsname
\providecommand{\newblock}{\relax}
\providecommand{\bibinfo}[2]{#2}
\providecommand{\BIBentrySTDinterwordspacing}{\spaceskip=0pt\relax}
\providecommand{\BIBentryALTinterwordstretchfactor}{4}
\providecommand{\BIBentryALTinterwordspacing}{\spaceskip=\fontdimen2\font plus
\BIBentryALTinterwordstretchfactor\fontdimen3\font minus \fontdimen4\font\relax}
\providecommand{\BIBforeignlanguage}[2]{{%
\expandafter\ifx\csname l@#1\endcsname\relax
\typeout{** WARNING: IEEEtran.bst: No hyphenation pattern has been}%
\typeout{** loaded for the language `#1'. Using the pattern for}%
\typeout{** the default language instead.}%
\else
\language=\csname l@#1\endcsname
\fi
#2}}
\providecommand{\BIBdecl}{\relax}
\BIBdecl

\bibitem{beshley2023emerging}
M.~Beshley, M.~Klymash, I.~Scherm \emph{et~al.}, ``Emerging network technologies for digital transformation: 5g/6g, iot, sdn/ibn, cloud computing, and blockchain,'' in \emph{Emerging Networking in the Digital Transformation Age}, M.~Klymash, A.~Luntovskyy, M.~Beshley, I.~Melnyk, and A.~Schill, Eds.\hskip 1em plus 0.5em minus 0.4em\relax Cham: Springer Nature Switzerland, 2023, pp. 1--20.

\bibitem{96_moussaoui2023divide}
M.~Moussaoui, E.~Bertin, and N.~Crespi, ``Divide and conquer: A business model agenda for beyond-5g and 6g,'' \emph{IEEE Communications Magazine}, vol.~61, no.~7, pp. 82--88, 2023.

\bibitem{dang2020should}
S.~Dang, O.~Amin, B.~Shihada \emph{et~al.}, ``What should 6g be?'' \emph{Nature Electronics}, vol.~3, no.~1, pp. 20--29, 2020.

\bibitem{shen2023five}
L.-H. Shen, K.-T. Feng, and L.~Hanzo, ``Five facets of 6g: Research challenges and opportunities,'' \emph{ACM Computing Surveys}, vol.~55, no.~11, pp. 1--39, 2023.

\bibitem{9114878}
R.~Adeogun, G.~Berardinelli, P.~E. Mogensen \emph{et~al.}, ``Towards 6g in-x subnetworks with sub-millisecond communication cycles and extreme reliability,'' \emph{IEEE Access}, vol.~8, pp. 110\,172--110\,188, 2020.

\bibitem{9083877}
G.~Berardinelli, P.~Mogensen, and R.~O. Adeogun, ``6g subnetworks for life-critical communication,'' in \emph{2020 2nd 6G Wireless Summit (6G SUMMIT)}, 2020, pp. 1--5.

\bibitem{9585402}
G.~Berardinelli, P.~Baracca, R.~O. Adeogun \emph{et~al.}, ``Extreme communication in 6g: Vision and challenges for `in-x' subnetworks,'' \emph{IEEE Open Journal of the Communications Society}, vol.~2, pp. 2516--2535, 2021.

\bibitem{10251878}
B.~Kim, D.~Calin, N.~Tenny \emph{et~al.}, ``Device centric distributed compute, orchestration and networking,'' \emph{IEEE Wireless Communications}, vol.~30, no.~4, pp. 6--8, 2023.

\bibitem{10428041}
M.~Dalgitsis, N.~Cadenelli, M.~A. Serrano \emph{et~al.}, ``Cloud-native orchestration framework for network slice federation across administrative domains in 5g/6g mobile networks,'' \emph{IEEE Transactions on Vehicular Technology}, pp. 1--14, 2024.

\bibitem{9648317}
H.~T. Nguyen, T.~Van~Do, and C.~Rotter, ``Scaling upf instances in 5g/6g core with deep reinforcement learning,'' \emph{IEEE Access}, vol.~9, pp. 165\,892--165\,906, 2021.

\bibitem{su20246g}
R.~Su, X.~Li, and D.~Taibi, ``6g software engineering: A systematic mapping study,'' \emph{arXiv preprint arXiv:2405.05017}, 2024.

\bibitem{yang2022liquid}
T.~Yang, M.~Qin, N.~Cheng \emph{et~al.}, ``Liquid software-based edge intelligence for future 6g networks,'' \emph{IEEE Network}, vol.~36, no.~1, pp. 69--75, 2022.

\bibitem{taleb20226g}
T.~Taleb, C.~Benza{\"\i}d, M.~B. Lopez \emph{et~al.}, ``6g system architecture: A service of services vision,'' \emph{ITU journal on future and evolving technologies}, vol.~3, no.~3, pp. 710--743, 2022.

\bibitem{alawadhi2022recent}
A.~Alawadhi and A.~Almogahed, ``Recent advances in edge computing for 6g,'' in \emph{2022 International Conference on Intelligent Technology, System and Service for Internet of Everything (ITSS-IoE)}.\hskip 1em plus 0.5em minus 0.4em\relax IEEE, 2022, pp. 1--6.

\bibitem{esposito2024extensive}
M.~Esposito, V.~Falaschi, and D.~Falessi, ``An extensive comparison of static application security testing tools,'' in \emph{Proceedings of the 28th International Conference on Evaluation and Assessment in Software Engineering}.\hskip 1em plus 0.5em minus 0.4em\relax ACM, 2024, pp. 69--78.

\bibitem{esposito2023can}
M.~Esposito, S.~Moreschini, V.~Lenarduzzi, D.~H{\"a}stbacka, and D.~Falessi, ``Can we trust the default vulnerabilities severity?'' in \emph{2023 IEEE 23rd International Working Conference on Source Code Analysis and Manipulation (SCAM)}.\hskip 1em plus 0.5em minus 0.4em\relax IEEE, 2023, pp. 265--270.

\bibitem{esposito2024correlation}
M.~Esposito, M.~Robredo, F.~A. Fontana, and V.~Lenarduzzi, ``On the correlation between architectural smells and static analysis warnings,'' \emph{arXiv preprint arXiv:2406.17354}, 2024.

\bibitem{DRS}
L.~Xiao, Y.~Cai, and R.~Kazman, ``Design rule spaces: a new form of architecture insight,'' in \emph{Proceedings of the 36th International Conference on Software Engineering}, ser. ICSE 2014.\hskip 1em plus 0.5em minus 0.4em\relax New York, NY, USA: Association for Computing Machinery, 2014, p. 967–977.

\bibitem{energyware}
Y.~Li, W.~Liang, J.~Li, X.~Cheng, D.~Yu, A.~Y. Zomaya, and S.~Guo, ``Energy-aware, device-to-device assisted federated learning in edge computing,'' \emph{IEEE TRANSACTIONS ON PARALLEL AND DISTRIBUTED SYSTEMS}, vol.~34, 2023.

\bibitem{wan2023software}
Z.~Wan, Y.~Zhang, X.~Xia, Y.~Jiang, and D.~Lo, ``Software architecture in practice: Challenges and opportunities,'' in \emph{Proceedings of the 31st ACM Joint European Software Engineering Conference and Symposium on the Foundations of Software Engineering}, 2023, pp. 1457--1469.

\bibitem{AI4RE}
A.~Mehraj, Z.~Zhang, and K.~Syst{\"a}, ``A tertiary study on~ai for~requirements engineering,'' D.~Mendez and A.~Moreira, Eds.\hskip 1em plus 0.5em minus 0.4em\relax Cham: Springer Nature Switzerland, 2024, pp. 159--177.

\bibitem{kotilainen2023towards}
P.~Kotilainen, V.~Heikkil{\"a}, K.~Syst{\"a} \emph{et~al.}, ``Towards liquid ai in iot with webassembly: A prototype implementation,'' in \emph{International Conference on Mobile Web and Intelligent Information Systems}.\hskip 1em plus 0.5em minus 0.4em\relax Springer, 2023, pp. 129--141.

\bibitem{agbese2023examining}
M.~Agbese, N.~M{\"a}kitalo, M.~Waseem \emph{et~al.}, ``Examining privacy and trust issues at the edge of isomorphic iot architectures: Case liquid ai,'' in \emph{Proceedings of the 13th International Conference on the Internet of Things}, 2023, pp. 245--252.

\bibitem{Arch}
P.~Kotilainen, A.~Mehraj, T.~Mikkonen \emph{et~al.}, ``The programmable world and its emerging privacy nightmare,'' in \emph{Proceedings of 24th International Conference on Web Engineering (ICWE)}.\hskip 1em plus 0.5em minus 0.4em\relax Springer Nature Switzerland, 2024.

\bibitem{bakhtin2023tools}
A.~Bakhtin, X.~Li, J.~Soldani \emph{et~al.}, ``Tools reconstructing microservice architecture: A systematic mapping study,'' \emph{Agility with Microservices Programming, co-located with ECSA}, vol. 2023, 2023.

\bibitem{cerny2022microservice}
T.~Cerny, A.~S. Abdelfattah, V.~Bushong \emph{et~al.}, ``Microservice architecture reconstruction and visualization techniques: A review,'' in \emph{2022 IEEE International Conference on Service-Oriented System Engineering (SOSE)}.\hskip 1em plus 0.5em minus 0.4em\relax IEEE, 2022, pp. 39--48.

\bibitem{daubaris2023explainability}
P.~Daubaris, S.~Linkola, J.~F. Ingl{\'e}s-Romero \emph{et~al.}, ``Explainability with observation sharing in long collaboration chains of automated systems-of-systems,'' \emph{IEEE Software}, 2023.

\bibitem{moreschini2023edge}
S.~Moreschini, E.~Younesian, D.~H{\"a}stbacka \emph{et~al.}, ``Edge to cloud tools: A multivocal literature review,'' \emph{Journal of Systems and Software}, p. 111942, 2023.

\bibitem{ESAAM23}
A.~Droob, D.~Morratz, F.~L. Jakobsen \emph{et~al.}, ``Flexconnect: Mobile computational offloading,'' in \emph{Proceedings of the 3rd Eclipse Security, AI, Architecture and Modelling Conference on Cloud to Edge Continuum}, ser. ESAAM '23, 2023, p. 29–38.

\bibitem{Fault}
------, ``Fault tolerant horizontal computation offloading,'' in \emph{2023 IEEE International Conference on Edge Computing and Communications (EDGE)}, 2023, pp. 177--182.

\bibitem{cerny2024static}
T.~Cerny, A.~S. Abdelfattah, J.~Yero \emph{et~al.}, ``From static code analysis to visual models of microservice architecture,'' \emph{Cluster Computing}, pp. 1--26, 2024.

\bibitem{thomsen2023edge}
J.~L. Thomsen, K.~D.~S. Thomsen, R.~B. Schmidt \emph{et~al.}, ``Edge computing tasks orchestration: An energy-aware approach,'' in \emph{2023 IEEE International Conference on Edge Computing and Communications (EDGE)}.\hskip 1em plus 0.5em minus 0.4em\relax IEEE, 2023, pp. 115--117.

\bibitem{kotilainen2022proposing}
P.~Kotilainen, T.~Autto, V.~J{\"a}rvinen \emph{et~al.}, ``Proposing isomorphic microservices based architecture for heterogeneous iot environments,'' in \emph{International Conference on Product-Focused Software Process Improvement}.\hskip 1em plus 0.5em minus 0.4em\relax Springer, 2022, pp. 621--627.

\bibitem{kotilainen2023webassembly}
P.~Kotilainen, V.~J{\"a}rvinen, J.~Tarkkanen \emph{et~al.}, ``{WebAssembly} in {IoT}: beyond toy examples,'' in \emph{International Conference on Web Engineering}.\hskip 1em plus 0.5em minus 0.4em\relax Springer, 2023, pp. 93--100.

\bibitem{schneider2024comparison}
S.~Schneider, A.~Bakhtin, X.~Li \emph{et~al.}, ``Comparison of static analysis architecture recovery tools for microservice applications,'' \emph{arXiv preprint arXiv:2403.06941}, 2024.

\bibitem{li2023evaluating}
X.~Li, D.~A. d'Aragona, and D.~Taibi, ``Evaluating microservice organizational coupling based on cross-service contribution,'' in \emph{International Conference on Product-Focused Software Process Improvement}.\hskip 1em plus 0.5em minus 0.4em\relax Springer, 2023, pp. 435--450.

\bibitem{cerny2023catalog}
T.~Cerny, A.~S. Abdelfattah, A.~Al~Maruf \emph{et~al.}, ``Catalog and detection techniques of microservice anti-patterns and bad smells: A tertiary study,'' \emph{Journal of Systems and Software}, vol. 206, p. 111829, 2023.

\bibitem{smith2023benchmarks}
S.~Smith, E.~Robinson, T.~Frederiksen \emph{et~al.}, ``Benchmarks for end-to-end microservices testing,'' in \emph{2023 IEEE International Conference on Service-Oriented System Engineering (SOSE)}.\hskip 1em plus 0.5em minus 0.4em\relax IEEE, 2023, pp. 60--66.

\bibitem{li2023metrics}
X.~Li, A.~S. Abdelfattah, R.~Su \emph{et~al.}, ``Metrics and models for developer collaboration analysis in microservice-based systems. a systematic mapping study,'' \emph{Joint Proceedings IWSM and MENSURA 2023}, 2023.

\bibitem{parker2023visualizing}
G.~Parker, S.~Kim, A.~Al~Maruf \emph{et~al.}, ``Visualizing anti-patterns in microservices at runtime: A systematic mapping study,'' \emph{IEEE Access}, vol.~11, pp. 4434--4442, 2023.

\bibitem{abdelfattah2023end}
A.~S. Abdelfattah, T.~Cerny, J.~Y. Salazar \emph{et~al.}, ``End-to-end test coverage metrics in microservice systems: An automated approach,'' in \emph{European Conference on Service-Oriented and Cloud Computing}.\hskip 1em plus 0.5em minus 0.4em\relax Springer, 2023, pp. 35--51.

\bibitem{yang2023exploring}
N.~Yang, ``Exploring the 6g software business ecosystem: A morphological analysis approach,'' \emph{Lappeenranta-Lahti University of Technology LUT Master's Thesis}, 2023.

\bibitem{akbar2024role}
M.~A. Akbar, A.~A. Khan, and S.~Hyrynsalmi, ``Role of quantum computing in shaping the future of 6 g technology,'' \emph{Information and Software Technology}, vol. 170, p. 107454, 2024.

\bibitem{akbar20246g}
M.~A. Akbar, A.~A. Khan, S.~Hyrynsalmi \emph{et~al.}, ``6g secure quantum communication: a success probability prediction model,'' \emph{Automated Software Engineering}, vol.~31, no.~1, p.~31, 2024.

\bibitem{esposito2024classi}
M.~Esposito, M.~T. Sabzevari, B.~Ye \emph{et~al.}, ``$classi|q\rangle$ towards a translation framework to bridge the classical-quantum programming gap,'' \emph{arXiv preprint arXiv:2406.06764}, 2024.

\bibitem{sabzevari2024qcshqd}
M.~T. Sabzevari, M.~Esposito, A.~A. Khan \emph{et~al.}, ``Qcshqd: Quantum computing as a service for hybrid classical-quantum software development: A vision,'' \emph{arXiv preprint arXiv:2403.08663}, 2024.

\bibitem{esposito2024beyond}
M.~Esposito, F.~Palagiano, and V.~Lenarduzzi, ``Beyond words: On large language models actionability in mission-critical risk analysis,'' \emph{arXiv preprint arXiv:2406.10273}, 2024.

\bibitem{esposito2024leveraging}
M.~Esposito and F.~Palagiano, ``Leveraging large language models for preliminary security risk analysis: A mission-critical case study,'' in \emph{Proceedings of the 28th International Conference on Evaluation and Assessment in Software Engineering}.\hskip 1em plus 0.5em minus 0.4em\relax ACM, 2024, pp. 442--445.

\bibitem{SP3}
N.~Katiyar, P.~Kumari, S.~Sakhshi \emph{et~al.}, ``Trending iot platforms on middleware layer,'' in \emph{Intelligent Analytics for Industry 4.0 Applications}.\hskip 1em plus 0.5em minus 0.4em\relax CRC Press, 2023, pp. 131--147.

\bibitem{SP7}
M.~A. Habibi, A.~G. S{\'a}nchez, I.~L. Pav{\'o}n \emph{et~al.}, ``Enabling network and service programmability in 6g mobile communication systems,'' in \emph{2022 IEEE Future Networks World Forum (FNWF)}.\hskip 1em plus 0.5em minus 0.4em\relax IEEE, 2022, pp. 320--327.

\bibitem{SP1}
A.~A. Ateya, A.~A. Alhussan, H.~A. Abdallah \emph{et~al.}, ``Edge computing platform with efficient migration scheme for 5g/6g networks.'' \emph{Computer Systems Science \& Engineering}, vol.~45, no.~2, 2023.

\bibitem{SP5}
G.~Manogaran, T.~Baabdullah, D.~B. Rawat \emph{et~al.}, ``Ai-assisted service virtualization and flow management framework for 6g-enabled cloud-software-defined network-based iot,'' \emph{IEEE Internet of Things Journal}, vol.~9, no.~16, pp. 14\,644--14\,654, 2021.

\bibitem{SP4}
N.~Meenakshi, M.~M. Jaber, R.~Pradhan \emph{et~al.}, ``Design systematic wireless inventory trackers with prolonged lifetime and low energy consumption in future 6g network,'' \emph{Wireless Networks}, pp. 1--22, 2023.

\bibitem{SP2}
I.~Al-Hammadi, M.~Li, and S.~M. Islam, ``Independent tasks scheduling of collaborative computation offloading for sdn-powered mec on 6g networks,'' \emph{Soft Computing}, vol.~27, no.~14, pp. 9593--9617, 2023.

\bibitem{SP11}
A.~Alotaibi and A.~Barnawi, ``Idsoft: A federated and softwarized intrusion detection framework for massive internet of things in 6g network,'' \emph{Journal of King Saud University-Computer and Information Sciences}, vol.~35, no.~6, p. 101575, 2023.

\bibitem{SP12}
A.~Shukla, N.~Ahmed, A.~Roy \emph{et~al.}, ``Softwarized management of 6g network for green internet of things,'' \emph{Computer Communications}, vol. 187, pp. 103--114, 2022.

\bibitem{SP13}
I.~H. Abdulqadder and S.~Zhou, ``Sliceblock: Context-aware authentication handover and secure network slicing using dag-blockchain in edge-assisted sdn/nfv-6g environment,'' \emph{IEEE Internet of Things Journal}, vol.~9, no.~18, pp. 18\,079--18\,097, 2022.

\bibitem{SP17}
N.~Janbi, I.~Katib, A.~Albeshri \emph{et~al.}, ``Distributed artificial intelligence-as-a-service (daiaas) for smarter ioe and 6g environments,'' \emph{Sensors}, vol.~20, no.~20, p. 5796, 2020.

\bibitem{SP8}
P.~D. Bojovi{\'c}, T.~Malba{\v{s}}i{\'c}, D.~Vujo{\v{s}}evi{\'c} \emph{et~al.}, ``Dynamic qos management for a flexible 5g/6g network core: A step toward a higher programmability,'' \emph{Sensors}, vol.~22, no.~8, p. 2849, 2022.

\bibitem{SP9}
J.~A. Alonso-Lupez, L.~A.~M. Hern{\'a}ndez, S.~P. Arteaga \emph{et~al.}, ``Level of trust and privacy management in 6g intent-based networks for vertical scenarios,'' in \emph{2022 1st International Conference on 6G Networking (6GNet)}.\hskip 1em plus 0.5em minus 0.4em\relax IEEE, 2022, pp. 1--4.

\bibitem{SP18}
Y.~Tao, J.~Wu, X.~Lin \emph{et~al.}, ``Drl-driven digital twin function virtualization for adaptive service response in 6g networks,'' \emph{IEEE Networking Letters}, 2023.

\bibitem{SP10}
H.~Cao, S.~Garg, G.~Kaddoum \emph{et~al.}, ``Softwarized resource management and allocation with autonomous awareness for 6g-enabled cooperative intelligent transportation systems,'' \emph{IEEE Transactions on Intelligent Transportation Systems}, vol.~23, no.~12, pp. 24\,662--24\,671, 2022.

\bibitem{SP14}
M.~Kamruzzaman and O.~Alruwaili, ``Ai-based computer vision using deep learning in 6g wireless networks,'' \emph{Computers and Electrical Engineering}, vol. 102, p. 108233, 2022.

\bibitem{SP6}
O.~O. Ajibola, T.~E. El-Gorashi, and J.~M. Elmirghani, ``Disaggregation for energy efficient fog in future 6g networks,'' \emph{IEEE Transactions on Green Communications and Networking}, vol.~6, no.~3, pp. 1697--1722, 2022.

\bibitem{SP15}
H.~Cao, J.~Du, H.~Zhao \emph{et~al.}, ``Toward tailored resource allocation of slices in 6g networks with softwarization and virtualization,'' \emph{IEEE Internet of Things Journal}, vol.~9, no.~9, pp. 6623--6637, 2021.

\bibitem{SP16}
T.~Ye, J.~Zhang, C.~Zhao \emph{et~al.}, ``Service function chain orchestration in 6g software defined satellite-ground integrated networks,'' in \emph{2022 6th International Conference on Communication and Information Systems (ICCIS)}.\hskip 1em plus 0.5em minus 0.4em\relax IEEE, 2022, pp. 71--76.

\bibitem{zhao2021natural}
L.~Zhao, W.~Alhoshan, A.~Ferrari \emph{et~al.}, ``Natural language processing for requirements engineering: A systematic mapping study,'' \emph{ACM Computing Surveys (CSUR)}, vol.~54, no.~3, pp. 1--41, 2021.

\bibitem{lyu2023systematic}
Y.~Lyu, H.~Cho, P.~Jung \emph{et~al.}, ``A systematic literature review of issue-based requirement traceability,'' \emph{Ieee Access}, vol.~11, pp. 13\,334--13\,348, 2023.

\bibitem{xu2023systematic}
C.~Xu, Y.~Li, B.~Wang \emph{et~al.}, ``A systematic mapping study on machine learning methodologies for requirements management,'' \emph{IET Software}, vol.~17, no.~4, pp. 405--423, 2023.

\bibitem{corral2022building}
A.~Corral, L.~E. S{\'a}nchez, and L.~Antonelli, ``Building an integrated requirements engineering process based on intelligent systems and semantic reasoning on the basis of a systematic analysis of existing proposals.'' \emph{JUCS: Journal of Universal Computer Science}, vol.~28, no.~11, 2022.

\bibitem{sonbol2022use}
R.~Sonbol, G.~Rebdawi, and N.~Ghneim, ``The use of nlp-based text representation techniques to support requirement engineering tasks: A systematic mapping review,'' \emph{Ieee Access}, vol.~10, pp. 62\,811--62\,830, 2022.

\bibitem{tao2024magis}
W.~Tao, Y.~Zhou, W.~Zhang \emph{et~al.}, ``Magis: Llm-based multi-agent framework for github issue resolution,'' \emph{arXiv preprint arXiv:2403.17927}, 2024.

\bibitem{zamani2021machine}
K.~Zamani, D.~Zowghi, and C.~Arora, ``Machine learning in requirements engineering: A mapping study,'' in \emph{2021 IEEE 29th International Requirements Engineering Conference Workshops (REW)}.\hskip 1em plus 0.5em minus 0.4em\relax IEEE, 2021, pp. 116--125.

\bibitem{sofian2022systematic}
H.~Sofian, N.~A.~M. Yunus, and R.~Ahmad, ``Systematic mapping: Artificial intelligence techniques in software engineering,'' \emph{IEEE Access}, vol.~10, pp. 51\,021--51\,040, 2022.

\bibitem{hou2024large}
X.~Hou, Y.~Zhao, Y.~Liu \emph{et~al.}, ``Large language models for software engineering: A systematic literature review,'' 2024.

\bibitem{aguilar2020systematic}
A.~R. Aguilar, J.~O. Ochar{\'a}n-Hern{\'a}ndez, and {\'A}.~J. S{\'a}nchez-Garc{\'\i}a, ``A systematic mapping study of artificial intelligence in software requirements.'' \emph{Res. Comput. Sci.}, vol. 149, no.~11, pp. 179--188, 2020.

\bibitem{lin2021traceability}
J.~Lin, Y.~Liu, Q.~Zeng \emph{et~al.}, ``Traceability transformed: Generating more accurate links with pre-trained bert models,'' in \emph{2021 IEEE/ACM 43rd International Conference on Software Engineering (ICSE)}.\hskip 1em plus 0.5em minus 0.4em\relax IEEE, 2021, pp. 324--335.

\bibitem{guo2017semantically}
J.~Guo, J.~Cheng, and J.~Cleland-Huang, ``Semantically enhanced software traceability using deep learning techniques,'' in \emph{2017 IEEE/ACM 39th International Conference on Software Engineering (ICSE)}.\hskip 1em plus 0.5em minus 0.4em\relax IEEE, 2017, pp. 3--14.

\bibitem{borg2017traceability}
M.~Borg, C.~Englund, and B.~Duran, ``Traceability and deep learning-safety-critical systems with traces ending in deep neural networks,'' \emph{Proc. of the Grand Challenges of Traceability: The Next Ten Years}, pp. 48--49, 2017.

\bibitem{bahyResourceUtilizationComparison2023}
M.~B. Bahy, N.~R. Dwi~Riyanto, M.~Z. Fawwaz Nuruddin~Siswantoro \emph{et~al.}, ``\BIBforeignlanguage{en}{Resource {Utilization} {Comparison} of {KubeEdge}, {K3s}, and {Nomad} for {Edge} {Computing}},'' in \emph{\BIBforeignlanguage{en}{2023 10th {International} {Conference} on {Electrical} {Engineering}, {Computer} {Science} and {Informatics} ({EECSI})}}.\hskip 1em plus 0.5em minus 0.4em\relax Palembang, Indonesia: IEEE, Sep. 2023, pp. 321--327.

\bibitem{telenykComparisonKubernetesKubernetesCompatible2021}
S.~Telenyk, O.~Sopov, E.~Zharikov \emph{et~al.}, ``\BIBforeignlanguage{en}{A {Comparison} of {Kubernetes} and {Kubernetes}-{Compatible} {Platforms}},'' in \emph{\BIBforeignlanguage{en}{2021 11th {IEEE} {International} {Conference} on {Intelligent} {Data} {Acquisition} and {Advanced} {Computing} {Systems}: {Technology} and {Applications} ({IDAACS})}}.\hskip 1em plus 0.5em minus 0.4em\relax Cracow, Poland: IEEE, Sep. 2021, pp. 313--317.

\bibitem{bohmProfilingLightweightContainer}
S.~Bohm and G.~Wirtz, ``\BIBforeignlanguage{en}{Profiling {Lightweight} {Container} {Platforms}: {MicroK8s} and {K3s} in {Comparison} to {Kubernetes}}.''

\bibitem{koziolekLightweightKubernetesDistributions2023}
H.~Koziolek and N.~Eskandani, ``\BIBforeignlanguage{en}{Lightweight {Kubernetes} {Distributions}: {A} {Performance} {Comparison} of {MicroK8s}, k3s, k0s, and {Microshift}},'' in \emph{\BIBforeignlanguage{en}{Proceedings of the 2023 {ACM}/{SPEC} {International} {Conference} on {Performance} {Engineering}}}.\hskip 1em plus 0.5em minus 0.4em\relax Coimbra Portugal: ACM, Apr. 2023, pp. 17--29.

\bibitem{energyware2}
A.~Zhu, S.~Guo, M.~Ma \emph{et~al.}, ``Computation offloading for workflow in mobile edge computing based on deep q-learning,'' in \emph{2019 28th Wireless and Optical Communications Conference (WOCC)}.\hskip 1em plus 0.5em minus 0.4em\relax IEEE, 2019, pp. 1--5.

\bibitem{energyware3}
K.~Peng, B.~Zhao, S.~Xue, and Q.~Huang, ``Energy- and resource-aware computation offloading for complex tasks in edge environment,'' \emph{Complexity}, 2020.

\bibitem{energyware4}
K.~Liao, J.~Yang, and L.~Miao, ``Mobile edge computing offload strategy based on energy aware,'' in \emph{International Conference on Network Communication and Information Security (ICNCIS 2021)}, vol. 12175.\hskip 1em plus 0.5em minus 0.4em\relax SPIE, 2022, pp. 223--230.

\bibitem{uusitalo2021hexa}
M.~A. Uusitalo, M.~Ericson, B.~Richerzhagen \emph{et~al.}, ``Hexa-x the european 6g flagship project,'' in \emph{2021 Joint European Conference on Networks and Communications \& 6G Summit (EuCNC/6G Summit)}.\hskip 1em plus 0.5em minus 0.4em\relax IEEE, 2021, pp. 580--585.

\bibitem{lv2020software}
Z.~Lv and N.~Kumar, ``Software defined solutions for sensors in 6g/ioe,'' \emph{Computer Communications}, vol. 153, pp. 42--47, 2020.

\bibitem{ClementsPaul1996ASoA}
P.~Clements, ``\BIBforeignlanguage{eng}{A survey of architecture description languages},'' in \emph{\BIBforeignlanguage{eng}{International Workshop on Software Specifications \& Design: Proceedings of the 8th International Workshop on Software Specification and Design; 22-23 Mar. 1996}}.\hskip 1em plus 0.5em minus 0.4em\relax IEEE Computer Society, 1996, pp. 16--25.

\bibitem{MISHRA200759}
P.~Mishra and N.~Dutt, ``Chapter 4 - architecture description languages,'' in \emph{Customizable Embedded Processors}, ser. Systems on Silicon, P.~Lenne and R.~Leupers, Eds.\hskip 1em plus 0.5em minus 0.4em\relax Burlington: Morgan Kaufmann, 2007, pp. 59--76.

\bibitem{haider2018ali}
U.~Haider, J.~D. McGregor, and R.~Bashroush, ``The ali architecture description language,'' \emph{ACM SIGSOFT Software Engineering Notes}, vol.~43, no.~4, pp. 52--52, 2018.

\bibitem{EgyhazyCsaba2007Cofa}
C.~Egyhazy, ``\BIBforeignlanguage{eng}{Comparison of five architecture description languages on design focus, security and style},'' in \emph{\BIBforeignlanguage{eng}{ICEIS 2007 - 9th International Conference on Enterprise Information Systems, Proceedings}}, vol. ISAS, 2007, pp. 270--277.

\bibitem{shaw1995abstractions}
M.~Shaw, R.~DeLine, D.~V. Klein \emph{et~al.}, ``Abstractions for software architecture and tools to support them,'' \emph{IEEE transactions on software engineering}, vol.~21, no.~4, pp. 314--335, 1995.

\bibitem{allen1997formal}
R.~Allen and D.~Garlan, ``A formal basis for architectural connection,'' \emph{ACM Transactions on Software Engineering and Methodology (TOSEM)}, vol.~6, no.~3, pp. 213--249, 1997.

\bibitem{garlan2010acme}
D.~Garlan, R.~Monroe, and D.~Wile, ``Acme: An architecture description interchange language,'' in \emph{CASCON First Decade High Impact Papers}, 2010, pp. 159--173.

\bibitem{luckham1995specification}
D.~C. Luckham, J.~J. Kenney, L.~M. Augustin \emph{et~al.}, ``Specification and analysis of system architecture using rapide,'' \emph{IEEE transactions on software engineering}, vol.~21, no.~4, pp. 336--354, 1995.

\bibitem{moriconi1997introduction}
M.~Moriconi and R.~A. Riemenschneider, ``Introduction to sadl 1.0: A language for specifying software architecture hierarchies,'' Citeseer, Tech. Rep., 1997.

\bibitem{medvidovic1996formal}
N.~Medvidovic, R.~N. Taylor, and E.~J. Whitehead~Jr, ``Formal modeling of software architectures at multiple levels of abstraction,'' \emph{ejw}, vol. 714, pp. 824--2776, 1996.

\bibitem{binns1996domain}
P.~Binns, M.~Englehart, M.~Jackson \emph{et~al.}, ``Domain-specific software architectures for guidance, navigation and control,'' \emph{International Journal of Software Engineering and Knowledge Engineering}, vol.~6, no.~02, pp. 201--227, 1996.

\bibitem{feiler2006architecture}
P.~H. Feiler, D.~P. Gluch, and J.~Hudak, ``The architecture analysis \& design language (aadl): An introduction,'' 2006.

\bibitem{DashofyEric2005Acaf}
E.~Dashofy, A.~Hoek, and R.~Taylor, ``\BIBforeignlanguage{eng}{A comprehensive approach for the development of modular software architecture description languages},'' \emph{\BIBforeignlanguage{eng}{ACM transactions on software engineering and methodology}}, vol.~14, no.~2, pp. 199--245, 2005.

\bibitem{WangZhuxiao2012AADL}
Z.~Wang, H.~Peng, J.~Guo \emph{et~al.}, ``\BIBforeignlanguage{eng}{An architecture description language based on dynamic description logics},'' in \emph{\BIBforeignlanguage{eng}{IFIP Advances in Information and Communication Technology}}, ser. IFIP Advances in Information and Communication Technology.\hskip 1em plus 0.5em minus 0.4em\relax Berlin, Heidelberg: Springer Berlin Heidelberg, 2012, vol. 385, pp. 157--166.

\bibitem{1604766}
J.~Revillard, S.~Cimpan, E.~Benoit \emph{et~al.}, ``Intelligent instrument design with archware adl,'' in \emph{Fourth Workshop on Model-Based Development of Computer-Based Systems and Third International Workshop on Model-Based Methodologies for Pervasive and Embedded Software (MBD-MOMPES'06)}, 2006, pp. 9 pp.--74.

\bibitem{CuenotPhilippe2010TEad}
P.~Cuenot, P.~Frey, R.~Johansson \emph{et~al.}, ``\BIBforeignlanguage{eng}{The east-adl architecture description language for automotive embedded software},'' in \emph{\BIBforeignlanguage{eng}{Lecture Notes in Computer Science (including subseries Lecture Notes in Artificial Intelligence and Lecture Notes in Bioinformatics)}}, vol. 6100, 2010, pp. 297--307.

\bibitem{BashroushRabih2006TMFA}
R.~Bashroush, I.~Spence, P.~Kilpatrick \emph{et~al.}, ``\BIBforeignlanguage{eng}{Towards more flexible architecture description languages for industrial applications},'' in \emph{\BIBforeignlanguage{eng}{Lecture Notes in Computer Science (including subseries Lecture Notes in Artificial Intelligence and Lecture Notes in Bioinformatics)}}, vol. 4344.\hskip 1em plus 0.5em minus 0.4em\relax Berlin, Heidelberg: Springer Berlin Heidelberg, 2006, pp. 212--219.

\bibitem{BashroushR.2008AAEA}
------, ``\BIBforeignlanguage{eng}{Ali: An extensible architecture description language for industrial applications},'' in \emph{\BIBforeignlanguage{eng}{15th Annual IEEE International Conference and Workshop on the Engineering of Computer Based Systems (ecbs 2008)}}.\hskip 1em plus 0.5em minus 0.4em\relax IEEE, 2008, pp. 297--304.

\bibitem{katz20186genesis}
M.~Katz, M.~Matinmikko-Blue, and M.~Latva-Aho, ``6genesis flagship program: Building the bridges towards 6g-enabled wireless smart society and ecosystem,'' in \emph{2018 IEEE 10th Latin-American Conference on Communications (LATINCOM)}.\hskip 1em plus 0.5em minus 0.4em\relax IEEE, 2018, pp. 1--9.

\bibitem{03_bhat2023ecosystem}
J.~Bhat and S.~Alqahtani, ``6g ecosystem: Current status and future perspective,'' \emph{IEEE Communications Standards Magazine}, 2023.

\bibitem{ahokangas2022envisioning}
P.~Ahokangas, M.~Matinmikko-Blue, and S.~Yrj{\"o}l{\"a}, ``Envisioning a future-proof global 6g from business, regulation, and technology perspectives,'' \emph{IEEE Communications Magazine}, vol.~61, no.~2, pp. 72--78, 2022.

\bibitem{12_azari2023thz}
M.~Azari, S.~Solanki, S.~Chatzinotas \emph{et~al.}, ``Thz-empowered uavs in 6g: Opportunities, challenges, and trade-offs,'' \emph{IEEE Transactions on Wireless Communications}, 2023.

\bibitem{liao2016ict}
H.~Liao, B.~Wang, B.~Li \emph{et~al.}, ``Ict as a general-purpose technology: The productivity of ict in the united states revisited,'' \emph{Information Economics and Policy}, vol.~36, pp. 10--25, 2016.

\bibitem{hyrynsalmi2019ecosystem}
S.~Hyrynsalmi and S.~M. Hyrynsalmi, ``Ecosystem: A zombie category?'' in \emph{2019 IEEE International Conference on Engineering, Technology and Innovation (ICE/ITMC)}.\hskip 1em plus 0.5em minus 0.4em\relax IEEE, 2019, pp. 1--8.

\bibitem{kovtun2023formalization}
V.~Kovtun, I.~Izonin, and M.~Gregus, ``Formalization of the metric of parameters for quality evaluation of the subject-system interaction session in the 5g-iot ecosystem,'' \emph{International Journal of Open Information Technologies}, 2023.

\end{thebibliography}
